\documentclass[pra,twocolumn,amsmath,amssymb,superscriptaddress,showpacs,longbibliography]{revtex4-1}
\usepackage{graphicx}
\usepackage{latexsym}
\usepackage{amsmath}
\usepackage{graphics}
\usepackage{amssymb}
\usepackage{layout}
\usepackage{verbatim}
\usepackage{amsfonts,epsfig}
\usepackage{natbib}
\usepackage{hyperref}

\DeclareMathOperator{\Tr}{Tr}

\newcommand{\aq}{\rho_{\Phi}}
\newcommand{\bq}{\rho_{\Psi}}
\newcommand{\xq}{\sigma_{\Psi}}
\newcommand{\yq}{\sigma_{\Phi}}

\begin{document}

\title{The relation between the quantum discord and quantum teleportation:
the physical interpretation of the transition point between different quantum discord
decay regimes}

\author{Katarzyna Roszak}
\affiliation{Institute of Physics, Wroc{\l}aw University of Technology,
50-370 Wroc{\l}aw, Poland}

\author{{\L}ukasz Cywi{\'n}ski}
\affiliation{Institute of Physics, Polish Academy of Sciences,
02-668 Warsaw, Poland}

\date{\today}

\begin{abstract}
We study quantum teleportation via Bell-diagonal mixed states of two qubits
in the context
of the intrinsic properties of the quantum discord. We show that when the quantum-correlated 
state of the two qubits is used for quantum teleportation
the character of the teleportation efficiency changes substantially depending on the
Bell-diagonal-state parameters, which can be seen when the worst-case-scenario or best-case-scenario
fidelity is studied. Depending on the parameter range, one of two types of single
qubit states is hardest/easiest to teleport. The transition between these two parameter
ranges coincides exactly with the transition between the range of classical correlation
decay and quantum correlation decay characteristic for the evolution of the quantum discord.
The correspondence provides a physical interpretation for the prominent feature
of the decay of the quantum discord.
\end{abstract}

\maketitle

\section{Introduction}
The quantum discord \cite{olliver01,henderson01,modi12}, a measure of bi- and multi-partite quantum correlations,
has attracted much attention recently.
This is due to the fact that the discord
indicates the existence of quantum correlations in many partially mixed states
which have no entanglement present. Specifically, the quantum discord, $\mathcal{D}$, does not display
any sudden death type phenomenon, since the zero-discord states form a set of measure zero \cite{ferraro10}. This 
subset of states 
is the set of all truly classical states chosen asymptotically by decoherence \cite{zurek03,Hornberger_LNP09}.
Hence, a smooth, continuous decoherence process 
cannot lead to a sudden and continued disappearance of quantum correlations
mid-evolution (before a fully mixed, completely dephased state is reached). This suggests
that the sudden death of entanglement signifies not, as was previously believed,
the disappearance of all quantum correlations, but the crossing of a threshold
of a given, small amount of correlations and the disappearance of quantum correlations of
a certain type. Although below this threshold many
quantum informational tasks are no longer possible, methods of performing
quantum computation on zero-entanglement states for which the quantum discord is non-zero
have already been devised \cite{meurer92,knill98,datta08,passante11,horodecki14}.
Furthermore, it has been very recently shown experimentally
how entanglement can be shared between distant parties via non-entangled states
with non-zero discord \cite{dakic12,silberhorn13,fedrizzi13,peuntinger13,vollmer13}.

Unfortunately, computing 
the quantum discord 
for an arbitrary density matrix
is an extremely involved task even in the simplest two-qubit case
\cite{huang14}. This led to
the emergence of the geometric quantum discord \cite{dakic10}, which is defined
as the minimal Hilbert-Schmidt distance of a given state from the set of zero-discord
states. Although an explicit formula for the geometric discord given a two-qubit density
matrix does not yet exist (one that does not require minimization over the set of
all zero-discord states), such formulas exist for the lower \cite{dakic10} 
and upper \cite{miranowicz12} bounds
on the geometric discord. 
The geometric discord is a good measure to distinguish
between zero-discord and non-zero-discord states, but because of the properties of
the Hilbert-Schmidt distance, it is not a good measure for the amount of quantum
correlations present in a given state. In fact, because the Hilbert-Schmidt distance
is sensitive to the purity of the state, the geometric discord may be increased
by non-unitary evolution of a single qubit (the unmeasured one) \cite{piani12,tufarelli12,hu12},
which should not increase inter-qubit quantum correlations. 
One solution to this problem is rescaling
the geometric discord following Ref.~\cite{tufarelli13}; 
a lot of work is also being done to find 
explicit formulas for discord measures utilizing more appropriate distance 
measures \cite{nakano12,paula13,aaronson13,spehner13}.

Regardless of these difficulties, there are properties of typical discord evolutions, $\mathcal{D}(t)$,
under decoherence processes that have already become apparent. Perhaps the most prominent
and most baffling is the transition between the regimes in which either the classical correlations, $\mathcal{C}$ (defined as the difference between the mutual information
and the quantum discord $\mathcal{D}$), 
or the quantum correlations (i.e.~$\mathcal{D}$ itself), exhibit a faster decay 
\cite{Maziero_PRA09,mazzola10,Lim_JPA14}. The most striking example of such a behavior was given in \cite{mazzola10}, where it was shown that for a certain initial Bell-diagonal state, the influence of phase-damping channel leads to the evolution, during which $\mathcal{D}$ remains constant for a finite time $t_{c}$ after the state initialization 
(while $\mathcal{C}(t)$ are decaying), and for longer times it decays (while the $\mathcal{C}(t)$ remain constant). In a more general case, the transition between the two regimes is manifested by discontinuity in the first derivative of $\mathcal{D}(t)$, i.e.~by a transition between two regimes governed by different decay functions. Such a behavior was also seen in the dynamics of the geometrical quantum discord
and rescaled geometric quantum discord \cite{tufarelli13}, and the transition was shown to occur at exactly the same point as the transition in the decay of $\mathcal{D}(t)$ \cite{roszak13,mazurek14}. The study of the geometric discord revealed similar transitions occurring for non-Bell-diagonal initial states and evolutions. 

The above discussion suggests that the transition between the ``classical'' and ``quantum'' decay regimes is a rather general feature of $\mathcal{D}$ dynamics. There is, however, no intuitive understanding of the possible operational meaning of this transition, which is not surprising in the light of the fact that the operational significance of the quantum discord 
is a subject of ongoing research \cite{Cavalcanti_PRA11,Gu_NP12}. In the following, we give a simple interpretation of an observable (and practically relevant) quantity that changes discontinuously at the transition point of discord dynamics. We investigate the quantum teleportation protocol with the entangled Bell state used for teleportation being subject to decoherence. It is well known that the time at which the \emph{average} teleportation fidelity becomes smaller than $2/3$ is the time at which the bipartite state used as a resource becomes separable 
\cite{Popescu_PRL94,horodecki99,verstraete03}. Here we focus on teleportation fidelity \emph{minimized over the teleported states} (i.e.~the worst-case-scenario fidelity)
or \emph{maximized over the teleported states} (i.e.~the best-case-scenario fidelity; the two situations
display an equivalent transition), and we show that at the time of the transition between the two regimes of 
quantum discord decay, the nature of the state for which teleportation has the lowest/highest fidelity changes.

\section{Classical and quantum decoherence regimes of quantum discord dynamics}
The quantum discord is defined as the difference 
between two classically equivalent formulas for mutual information \cite{olliver01}. 
The formula which is referred to as mutual information in Ref.~\cite{mazzola10}
is given by \cite{olliver01}
\begin{equation}
\label{mi1}
\mathcal{I}(\rho_{AB})=S(\Tr_A\rho_{AB})+S(\Tr_B\rho_{AB})-S(\rho_{AB}),
\end{equation}
where the von Neumann entropy is given by $S(\rho)=-\Tr \rho\log_2\rho$.
This quantity was generalized from the classical language of probability distributions
in a straightforward manner to the language of density matrices, while the Shannon entropy
was replaced by von Neumann entropy.
The other formula for classical mutual information, which is in the quantum context 
referred to as classical correlations
(in Ref.~\cite{mazzola10}), cannot be generalized in a direct manner,
because the classical formula involves conditional entropy,
\begin{equation}
\nonumber
\mathcal{C}(A:B)=H(A)-H(A|B),
\end{equation}
where $H$ denotes the Shannon entropy and $A$ and $B$ are random variables.
Conditional entropy $H(A|B)$ requires the specification of the state of $A$ given the state of $B$,
which in quantum mechanics is ambiguous until the measurement performed on $B$ is specified.
Hence, the conditional von Neumann entropy can be found given the complete measurement on subsystem $B$
and the resulting formula for classical correlations is \cite{olliver01,henderson01}
\begin{equation}
\label{mi2}
\mathcal{C}(\rho_{AB})=\max_{\{\Pi_k\}}\left[S(\Tr\rho_{AB})-S(\rho_{AB}|\{\Pi_k\})\right],
\end{equation}
where $\{\Pi_k\}$ is a complete set of orthonormal projective operators corresponding 
to a von Neumann measurement of subsystem $B$. The index $k$ denotes the outcome of a given measurement
and the formula involves maximization over the set of projective measurements.
Therefore the formula of Eq.~(\ref{mi2})
yields the information gained about the system $A$ after the measurement $\{\Pi_k\}$ on system $B$.
The quantum discord of a given state $\rho_{AB}$ is then given by
\begin{equation}
\label{disc}
\mathcal{D}(\rho_{AB})=\mathcal{I}(\rho_{AB})-\mathcal{C}(\rho_{AB}).
\end{equation}

Following Ref.~\cite{mazzola10}, the value of the quantum discord can be found
for any Bell-diagonal two-qubit state.
If the Bell-diagonal two-qubit density matrix is written in the form
\begin{equation}
\label{X}
\rho_{AB}=\left(
\begin{array}{cccc}
\aq&0&0&\yq\\
0&\bq&\xq&0\\
0&\xq&\bq&0\\
\yq&0&0&\aq
\end{array}
\right),
\end{equation}
where all four parameters are real, 
the mutual information is given by 
\begin{eqnarray}
\nonumber
\mathcal{I}(\rho_{AB})&=&2+\sum_{i=\Psi,\Phi}\left[
(\rho_i+\sigma_i)\log_2(\rho_i+\sigma_i)\right.\\
\label{mib}
&&\left. +(\rho_i-\sigma_i)\log_2(\rho_i-\sigma_i)
\right].   
\end{eqnarray}
Note that any continuous and differentiable evolution of $\rho_{AB}$ that retains the Bell-diagonal form
of Eq.~(\ref{X}) must lead to a continuous and differentiable evolution of $\mathcal{I}(\rho_{AB})$.
An explicit formula for the classical correlations of Eq.~(\ref{mi2}) can also be found for any Bell-diagonal
state \cite{Maziero_PRA09}. These correlations are given by
\begin{equation}
\label{ccb}
\mathcal{C}(\rho_{AB})=\frac{1}{2}\left[(1+\chi)\log_2(1+\chi)
+(1-\chi)\log_2(1-\chi)\right],
\end{equation}
where $\chi=\max\{|\Delta|,|\xq|+|\yq|\}$, in which we used $\Delta \! \equiv \! \aq-\bq$.
The maximization allows for indifferentiability points in the evolution of classical correlations
which occurs when the plane 
\begin{equation}
\label{warunek}
|\Delta|=|\sigma_{\Psi}|+|\sigma_{\Phi}|
\end{equation}
is transgressed, resulting in indifferentiability points of the quantum discord
which is given by the difference of the smooth mutual information and the indifferentiable
classical correlation function, Eq.~(\ref{disc}).
Note that under a pure dephasing decoherence process, the classical correlation function remains constant
in the $|\Delta|\! > \! |\xq|+|\yq|$ regime, which we will call the {\it quantum decoherence regime}
following Ref.~\cite{mazzola10}, while it decays in the {\it classical decoherence regime},
$|\Delta|\! < \! |\xq|+|\yq|$. 
Below we consider decoherence which leads to $\mathcal{C}(t)$ and $\mathcal{D}(t)$ both decaying in the two regimes defined by the above inequalities.

\section{Relation between teleportation fidelity and discord}
To gain some understanding of the physical significance of this transition,
let us turn to the quantum teleportation of an unknown qubit state by means of
an entangled two-qubit state \cite{Bennett_PRL93} in the situation when the entangled state
has previously undergone partial decoherence. We will focus on the scenario where the
initial entangled state is the $|\Phi^+\rangle =1/\sqrt{2}(|00\rangle+|11\rangle)$
Bell state,
corresponding to $\aq=\yq=1/2$ and $\bq=\xq=0$
in Eq.~(\ref{X});
the assumption is only made for simplicity; the same results are
acquired regardless of the chosen Bell state. For such an initial state, decoherence
often retains the Bell-diagonal form, 
but rarely provides the means to induce
coherences between the $|01\rangle$ and $|10\rangle$ components
($\xq\! \neq \! 0$). It can, however,
induce non-zero occupations corresponding to $|01\rangle$ and $|10\rangle$
\cite{Aolita_RPP15,Cywinski_APPA11}, and possibly change the diagonal matrix elements in such a way that the state ceases to be Bell-diagonal. Since we want to maintain the Bell-diagonal form at all times (so that we can analytically calculate the quantum discord), below we will focus on a physically well motivated example of decohering channel which preserves this form with $\xq\! = \! 0$.

First let us explain the connection between the two regimes of discord dynamics and the teleportation fidelity. 
The unknown qubit to be teleported is $|\psi\rangle =\alpha|0\rangle +\beta|1\rangle$,
with $|\alpha|^2+|\beta|^2=1$.
After teleporting $|\psi\rangle$ using the state of Eq.~(\ref{X}) with $\sigma_{\Psi}=0$
instead of the $|\Phi^+\rangle$, the teleported state is equal to
\begin{eqnarray}
\label{tel}
\rho_{t}&=&|0\rangle\langle 0| (2\aq|\alpha|^2+2\bq|\beta|^2)\\
\nonumber
&&+|1\rangle\langle 1| (2\bq|\alpha|^2+2\aq|\beta|^2)+|0\rangle\langle 1| 2\yq\alpha\beta^* +\mathrm{H.c..} \,\, ,
\end{eqnarray}
and the fidelity of teleportation is given by 
\begin{equation}
\label{fid1}
F^2=\langle \psi|\rho_{t}|\psi\rangle 
=2\aq - 4(\Delta-\yq)|\alpha|^2(1-|\alpha|^2) \,\, .  
\end{equation}
The standard procedure now is to find the average fidelity of the teleported state
(averaged over all possible states to be teleported) and 
compare it to the classical limit of teleportation capability which is equal to $2/3$ \cite{Popescu_PRL94,horodecki99,verstraete03}. The average fidelity is given by
\begin{equation}
F^{2}_{\text{av}} = 2\aq - \frac{2}{3}(\Delta-\yq) \,\, , \label{eq:Fav}
\end{equation}
which is a smooth function of time.
In order to see the correspondence between the transition in the evolution of 
quantum correlations and teleportation capability, one should rather look at the 
extremal scenarios, namely, the fidelity of the teleported state minimized or maximized over all the $|\psi\rangle$ states. 
Analyzing the extrema of Eq.~(\ref{fid1}) as a function of $|\alpha|^2\in[0,1]$ we find that for $|\Delta|<|\yq|$ (in the classical decoherence regime), the minimal fidelity $F^2_{\min}=1/2+\Delta$ occurs for $|\psi\rangle = |0\rangle, |1\rangle$ (the poles on the Bloch sphere), and the maximal fidelity $F^2_{\text{max}}=1/2+\yq$ is obtained for $|\psi\rangle = 1/\sqrt{2} (|0\rangle +\exp(i\phi)|1\rangle$ (the states on the equator of the Bloch sphere). On the other hand, for $|\Delta|>|\yq|$ (in the quantum decoherence regime), the states which are easiest/harder to teleport (with the corresponding fidelities) trade places, so that we always have $F^2_{\min}=1/2+\text{min}(\Delta,\yq)$ and $F^2_{\text{max}}=1/2+\text{max}(\Delta,\yq)$.
At the transition point, $\Delta \! = \!\yq$, the fidelity from Eq.~(\ref{fid1}) is independent of the state $|\psi\rangle$, so that it is equal to its average value, $2\aq \! = \! \Delta + 1/2$. If $\Delta \! > \! 1/6$ at this point in time, the transition between the two regimes of quantum discord dynamics 
occurs when the state is still entangled, and $F^{2}_{\text{av}}$ is larger than its maximal classical value.

The character of state $|\phi\rangle$ which is easiest/hardest to teleport in a given regime can be also connected with the character of the classical states closest to $\rho_{AB}$ \cite{Modi_PRL10}. In the ``classical'' regime there are two such states, $\rho_{x}$ and $\rho_{y}$ given by
\begin{eqnarray}
\rho_{a} & = & \left(\frac{1}{4}+\frac{\yq}{2} \right)(|aa\rangle\langle aa| + |\bar{a}\bar{a}\rangle\langle \bar{a}\bar{a}|) + \nonumber \\
& & \left(\frac{1}{4}-\frac{\yq}{2} \right)(|a\bar{a}\rangle\langle a\bar{a}| + |\bar{a}a\rangle\langle \bar{a}a|) \label{eq:ccs}
\end{eqnarray}
where $a\! = \! x,y$, and the states $|a\rangle$ and $|\bar{a}\rangle$ are eigenstates of
the Pauli matrices $\sigma_{a}$ with $\pm \! 1$ eigenvalues. Such a classically correlated state $\rho_{a}$ allows for ``teleportation'' of $|\phi\rangle \! =\! |a\rangle$, $|\bar{a}\rangle$ with fidelity of $1/2+\yq$. The fact that these are the closest classical states in $\yq\!>\! \Delta$ regime gives an intuitive explanation for the maximal fidelity being achieved for $|\phi\rangle$ state having its Bloch vector in the $xy$ plane. On the other hand, in the ``quantum'' regime the closest classical state is given by formula analogous to the one from Eq.~(\ref{eq:ccs}), but with $a\! = \! z$ and $\yq$ being replaced by $\Delta$. The correlations in this state allow for classical enhancement of teleportation fidelity of $|0\rangle$ and $|1\rangle$ states. 

\begin{figure}
\includegraphics[scale=0.7]{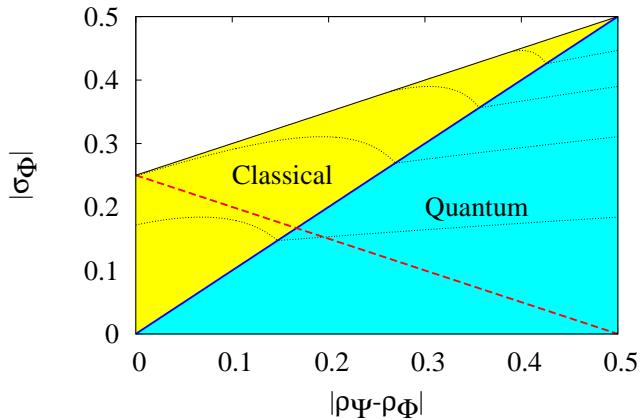}
\caption{\label{fig:mapka} Regimes of decoherence depending on the Bell-diagonal density matrix 
parameters for $\sigma_{\Psi}=0$. The blue line denotes the indifferentiability plane
given by Eq.~(\ref{warunek}),
dividing the quantum decoherence regime (cyan triangle) from the classical decoherence regime
(yellow triangle). The uncolored region corresponds to parameter values that yield
an unphysical density matrix. The red dashed line is the border, zero-entanglement line
(entanglement vanishes below the line). The dotted lines are izodiscords, corresponding to 
$D=0.1, 0.3, 0.5, 0.7$ going from bottom to top.} 
\end{figure}

Fig.~\ref{fig:mapka} shows the regimes of decoherence for Bell-diagonal density matrices with
$\xq=0$ with respect to the amplitude of the coherence present in the system
and the difference of the two distinct two-qubit occupations. 
The border between the region of quantum decoherence (cyan) and classical
decoherence (yellow) in terms of the quantum discord is denoted by the blue line.
The Bell state used above is located in the upper right corner of the figure 
($\Delta=1/2$ and $|\sigma_{\Phi}|=1/2$)
and its decoherence will move it to the lower left. In fact, a pure dephasing
process of an initial Bell state results in its moving straight down along the 
$\Delta=1/2$ line, which is located solely in the quantum decoherence regime. 
Vanishing entanglement corresponds to the point when
$F^{2}_{\text{av}}$  reaches the classical communication limit of $2/3$ \cite{Popescu_PRL94,horodecki99,verstraete03}. This boundary is denoted by a red, dashed line in Fig.~\ref{fig:mapka}. As seen, 
sudden death of entanglement \cite{zyczkowski01,yu04,eberly07} (i.e.~crossing of the separability boundary while $\yq$ is nonzero)  can be achieved both via the quantum decoherence regime and the classical decoherence regime. Moreover, entanglement sudden death can occur for higher values of coherence $\yq$ in the ``classical'' parameter range, because it is susceptible to the disturbance of qubit occupations \cite{roszak06a}.
Incidentally, the zero-discord line is the $|\sigma_{\Phi}|=0$ axis and coincides with 
the minimized teleportation fidelity reaching the minimal value  of $1/2$.

It is straightforward to generalize the teleportation procedure via a decohered two-qubit state
to any Bell-diagonal state, Eq.~(\ref{X}), and to show that the discord transition plane of 
Eq.~(\ref{warunek}) still
corresponds to a transition in the worst-case-scenario teleportation from the situation when the
states on the pole of the Bloch sphere are hardest to teleport, and when states on the equator of
the Bloch sphere are hardest to teleport.
Hence the relation between quantum teleportation and the quantum
discord presented in this Section holds for any state of the form given
by Eq.~(\ref{X}).

\section{Possible physical realization}
Let us now present a model of open-system dynamics in which the above-discussed transition between two teleportation regimes occurs at finite time $t_{c}$. It is clear that the classical decoherence requires a minimal amount of dephasing (decay of $\yq$) consistent with a given amount of relaxation (decay of $\Delta$), suggesting the use of amplitude damping channel. In order for the state to remain Bell-diagonal, we use the generalized amplitude damping channel \cite{Aolita_RPP15} corresponding to an environment being in equilibrium at temperature much higher than the energy splitting of the qubits. Assuming Markovian decoherence, we have then $\Delta(t) \! = \! \Delta(0)e^{-\Gamma t}$ and $\yq(t) \! = \! \yq(0)e^{-\Gamma t/2}$, where $\Gamma \! =\! \Gamma_{A}+\Gamma_{B}$, with $\Gamma_{A,B}$ being the longitudinal relaxation rates of the two qubits forming the entangled state used for teleportation.

However, under the action of this channel alone, the evolution starting form one of the Bell states remains in the classical regime at all times. In order for the transition to the quantum decoherence regime to occur, the additional dephasing process, having negligible contribution at short times, has to become stronger at longer times. This will happen when the qubits forming the entangled state used for teleportation are
coupled via their $\hat{\sigma}_{z}$ operators to a slowly fluctuating bath. Low frequency ($\omega \! \ll \! k_{B}T$, where $T$ is the bath temperature) environmental fluctuations can be treated as classical noise $\xi(t)$ \cite{Schoelkopf_spectrometer}, which under a realistic assumption of Gaussian statistics is described by a spectral density $S(\omega) \! \equiv \! \int e^{i\omega t} \langle\xi(t)\xi(0)\rangle \text{d}t$ (where $\langle ... \rangle$ denotes averaging over noise realizations. When most of spectral weight is concentrated at low frequencies ($\omega \! \ll \! 1/T_{2}$ where $T_{2}$ is the characteristic dephasing timescale), the total noise power is $\sigma^{2} \! \propto \! \int_{0}^{1/T_{2}} S(\omega) \text{d}\omega$. This is a typical situation in many solid-state based qubits (with $1/f$ charge and flux noise \cite{Paladino_RMP14} and hyperfine interaction with nuclear bath \cite{Hanson_RMP07,Fischer_SSC09,Cywinski_APPA11} being prominent examples). Such a noise leads to an additional decay of $\yq(t)$, given by $\yq(t) \rightarrow \yq(t)\exp[-(\gamma t)^2]$, with $\gamma \! =\! \sigma/\sqrt{2}$. Note that since the dephasing and relaxation are caused by very different processes (the former by low-energy fluctuations in the environment, the latter by high-energy processes involving exchanges of energy quanta corresponding to qubits' energy splittings, which are typically $\gg \! k_{B}T$), treating them as independent and additive processes is a reasonable approximation.
Decay of quantum correlations due to such low-frequency or quasi-static baths was a subject of a few recent works \cite{Bellomo_PRA10,benedetti13,rossi14,mazurek14,Mazurek_PRA14,Bragar_PRB15}. 
Within this model of decoherence we obtain a transition between two regimes of quantum discord decay at $t_{c} \! = \! \Gamma/2\gamma^{2}$, at which point we have $F^{2}_{\text{min}} \! = \! \frac{1}{2}[1+\exp(-\frac{1}{2}(\Gamma/\gamma)^{2}]$, so that this worst-case scenario fidelity still exhibits entanglement-related enhancement when $\Gamma/\gamma \! < \! \sqrt{2\ln 3}$. The modulus of the discontinuity of the derivative of $\mathcal{D}(t)$ at $t\!=\! t_{c}$ is given by $\frac{1}{2}\Gamma\Delta_{c} \ln(\frac{1+2\Delta_{c}}{1-2\Delta_{c}})$, where $\Delta_{c}\! = \! \frac{1}{2}\exp(-\Gamma/\gamma)$. The quantum-classical transition is thus most visible for low values of $\Gamma/\gamma$, which means that the rms $\sigma$ of the low-frequency phase noise should be much larger than the energy relaxation rate of the qubits. When the energy relaxation is in fact caused by transverse coupling to high-frequency noise, the power of this noise at frequencies corresponding to qubits' energy splittings (which is proportional to $\Gamma$) should be much smaller than $\sigma$. When the low- and high-frequency noises have the same  physical origin, this requirement is consistent with our assumption of the low-frequency character of the environmental fluctuations.

Note that while the above result has been obtained in the case of pure dephasing coupling to low-frequency noise (which leads to Gaussian decay of $\yq(t)$), a qualitatively analogous transition between classical and quantum decoherence regimes can occur for more general coupling to low-frequency noise. For example, for transverse coupling to noise in the presence of large energy splitting (i.e.~the qubit Hamiltonian $\hat{H} \! =\!\Omega \hat{\sigma}_{z}/2 + \xi(t)\hat{\sigma}_{x}/2$ with $\Omega \! \gg \! \sqrt{\langle\xi^{2}\rangle}$), the evolution of the qubits is approximately \cite{Makhlin_PRL04,Cucchietti_PRA05,benedetti14,Cywinski_PRA14,Szankowski_QIP15} of the pure dephasing form (only with the term quadratic in the noise, $\xi^{2}(t)/2\Omega$, contributing to the qubits' splittings), and the initial coherence decay is $\yq(t) \! \approx \! 1- (t/\gamma')^{2}$, with $\gamma' \! \approx \! \gamma^{2}/\Omega$. At very short times $ (t/\gamma')^{2} \! \ll \! \Gamma t$, and the discord dynamics will be in the classical regime, while at longer times the quantum decoherence regime can be entered. Interestingly, since the asymptotic decay of $\yq(t)$ in this case is of power-law kind \cite{Makhlin_PRL04,Bellomo_PRA10,Szankowski_QIP15} (so that for long $t$ again we have $\yq(t) \! > \! \Delta(t)$), the transition into the quantum decoherence regime has to be followed by a re-entry into the classical decoherence regime at a later time.

\section{Conclusion}
Summarizing, we have studied the properties of quantum correlations present in mixed two-qubit states
of Bell-diagonal form.
We have shown that there is a direct correspondence between the transition point
between the regime of classical correlation decay and the regime of quantum correlation
decay, which is a characteristic of the evolution of the quantum discord, and the
transition point which appears for the worst-case-scenario Fidelity of quantum teleportation.
For teleportation, the two regions being transgressed correspond
to two different classes of states which are hardest to teleport. Those are either
the equal superposition states in the quantum decoherence regime, or the single-component 
$|0\rangle/|1\rangle$ states in the classical decoherence regime. This shows that there
is a qualitative physical difference between the quantumly correlated states studied,
depending on the correlation decay regime in which the state is located.
Furthermore, the fact that the transition point is seen also by studying the geometric discord 
and that similar transition points are seen for initial states which
for technical reasons cannot be 
handled using the regular quantum discord, suggests that there is an underlying
physical meaning to all such transition points, and that classical-quantum decoherence transition
is a widely occurring property of the quantum discord.

\acknowledgments
K.R.~acknowledges support from the
National Science Centre project 2011/01/B/ST2/05459
and {\L}.C.~acknowledges support from the
National Science Centre project DEC-2012/07/B/ST3/03616.


%

\end{document}